\documentclass[times]{aastex631}
\usepackage{comment}

\emergencystretch=\maxdimen
\begin{document}
\title{\textsc{ArchGEM}: an Advanced Data Analysis Tool for Analyzing Scattered Light Noise in LIGO} 

\author[0009-0009-5018-848X]{Kaylah McGowan}
\affiliation{Department of Physics and Astronomy, Vanderbilt University\\ 2201 West End Avenue, Nashville, Tennessee, 37235}

\author[0009-0001-1750-3531]{Shania Nichols}
\affiliation{Department of Physics and Astronomy, Louisiana State University\\ 202 Nicholson Hall, Baton Rouge, Louisiana, 70803}
\affiliation{SETI Institute, NANOGrav Collaboration\\ 339 N Bernardo Ave Suite 200, Mountain View, California, 94043}
\author[0000-0003-3856-8534]{Siddharth Soni}
\affiliation{Massachusetts Institute of Technology, McNair Building\\
70 Vassar St, Cambridge, Massachusetts, 02139}
\author[0000-0001-8700-3455]{Chayan Chatterjee}
\affiliation{Department of Physics and Astronomy, Vanderbilt University\\ 2201 West End Avenue, Nashville, Tennessee, 37235}
\affiliation{Data Science Institute, Vanderbilt University\\ 1400 18th Avenue South Building, Suite 2000, Nashville, Tennessee, 37212}

\author[0000-0003-0199-3158]{Gabriela Gonzalez}
\affiliation{Department of Physics and Astronomy, Louisiana State University\\ 202 Nicholson Hall, Baton Rouge, Louisiana, 70803}

\author[0000-0003-2227-1322]{Kelly Holley--Bockelmann}
\affiliation{Department of Physics and Astronomy, Vanderbilt University\\ 2201 West End Avenue, Nashville, Tennessee, 37235}

\author[0000-0003-1007-8912]{Karan Jani}
\affiliation{Department of Physics and Astronomy, Vanderbilt University\\ 2201 West End Avenue, Nashville, Tennessee, 37235}



\begin{abstract}
Scattered light is one of the most common sources of non-stationary noise at low frequencies in Advanced LIGO detectors. It appears as arch-like features in time-frequency spectrograms, produced when stray light reflects from moving surfaces and recombines with the main interferometer beam. In this study, we present \textsc{ArchGEM}, an automated framework for identifying and characterizing these arches and recovering the physical properties of the scattering surfaces. \textsc{ArchGEM} combines a prominence-based peak-finding method with a Gaussian Mixture Model clustering approach to capture a range of scattered-light morphologies across different detector conditions. We apply \textsc{ArchGEM} to scattered light glitches across Advanced LIGO observing runs O3 (2019--2020) and O4 (2023--2024). We find that the average frequency distributions of these noise span $15-25$ Hz in O3a and O4 but increased to $20-40$ Hz during O3b. Typical inferred surface velocities are $0.2-0.5~\mu m/s$, and inferred surface displacements are $0.1-0.3~\mu m$. The Gaussian Mixture Model performs most consistently for complex or overlapping features, with mean frequency offsets within 5 Hz of the Gravity Spy baseline. Our results show that \textsc{ArchGEM} provides a practical tool for detector characterization by linking observed spectrogram features to the motion of scattering surfaces and helping guide future mitigation of scattered light noise in current and next-generation interferometers.
By quantifying the temporal and spectral behavior of scattered light, \textsc{ArchGEM} provides a robust framework for diagnosing noise sources and guiding targeted mitigation strategies in future detector upgrades.
\end{abstract}

\section{Introduction}
\label{sec:intro}

The Advanced LIGO detectors, together with Advanced Virgo and KAGRA as part of the LIGO--Virgo--KAGRA (LVK) Collaboration, have enabled direct observations of gravitational waves, providing unprecedented insights into compact binary mergers and relativistic astrophysics \citep{Abbott2016,Abbott2019}. These detections have not only validated Einstein’s theory of general relativity \citep{Einstein_GR} but also allowed precise tests of gravity in the strong-field regime and constraints on the neutron-star equation of state \citep{GW170817_EOS}. Achieving the required detector sensitivity demands a continual battle against a broad range of instrumental and environmental noise sources \citep{LIGO_India}. During the fourth observing run (O4), commissioning and detector-characterization work enabled the LIGO Hanford and Livingston detectors to reach their best sensitivities to date \citep{Capote2025O4perf,Soni2024O4DetChar}. Among these, scattered light remains one of the most frequent and persistent classes of transient noise that affects the LIGO low-frequency regime (10--60~Hz) \citep{Soni2021,Soni2024}.  

\begin{figure*}[t!]
    \centering \includegraphics[width=0.7\textwidth]{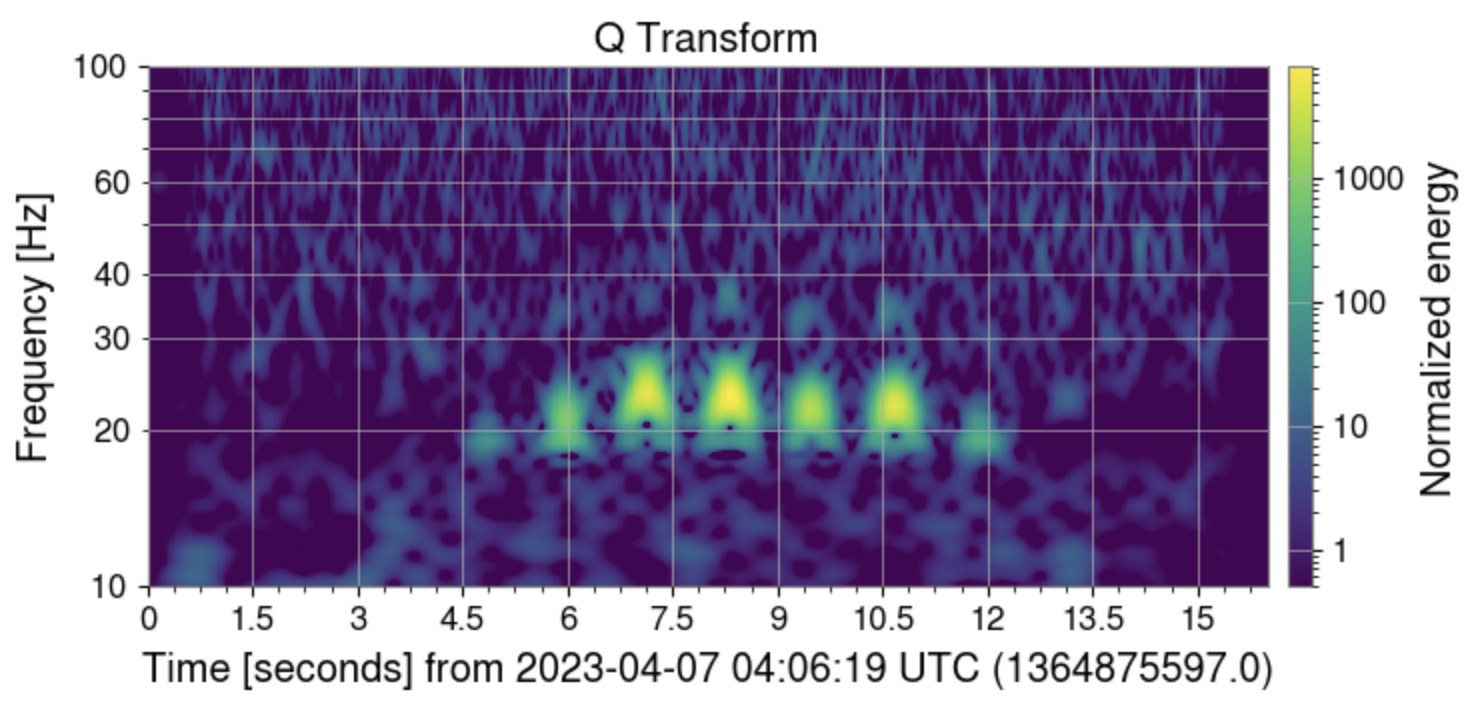}
\caption{Q-scan of scattered light noise observed in the calibration strain channel at LIGO Livingston during the Third Observing Run (O3). The plot highlights the distinct arch-like morphology of the scattered light noise, indicating periodic motion of a scattering surface within the interferometer environment. This noise predominantly affects the gravitational-wave data between 20–50 Hz.}
\label{fig:scatter}
\end{figure*}

Scattered-light glitches originate when stray laser light reflects off moving surfaces such as optics, baffles, or vacuum chamber walls and recombines with the main interferometer beam. The relative motion of the scattering surface introduces time-varying phase noise that manifests as characteristic arches in the time--frequency representation of the gravitational-wave strain channel, as shown in Figure~\ref{fig:scatter}. These events are unpredictable, non-stationary, and display highly variable morphologies depending on the amplitude and frequency of surface motion, microseismic conditions, and local environmental disturbances \citep{Ottaway2012, Nuttall2018}. Because scattered light is an inherent artifact of the interferometer’s optical design, it cannot be completely eliminated and will remain a recurring source of contamination for current and future detectors. Scattered light arises when stray reflections modulate with moving optical surfaces, often enhanced by residual ground motion. The Advanced LIGO isolation strategy mitigates these disturbances through multiple active inertial stages and hydraulic pre-isolators \citep{seismic_noise_3}. Despite these controls, elevated ground motion remains a dominant precursor to scattered-light noise, motivating targeted veto-generation methods \citep{seismic_noise_1}. Within the overall detector noise budget \citep{LIGO_detector_diagram}, scattered light represents one of the most non-Gaussian, environment-coupled components.  

Scattered-light noise is strongly driven by microseismic motion, which dominates ground displacement in the 0.1–0.3 Hz band and couples nonlinearly into the interferometer optics. Such ground motion, arising from oceanic wave activity or local anthropogenic disturbances, excites suspended components and optical tables, thereby modulating the relative motion between scattering surfaces and the main beam \citep{seismic_noise_2,seismic_noise_3}. During periods of elevated microseism, these interactions amplify the phase noise of the scattered field, shifting its spectral power into the gravitational-wave band (10–60 Hz) and producing the characteristic arch-like patterns observed in the strain spectrograms. This coupling mechanism underscores why scattered light is particularly prevalent at low frequencies and remains one of the dominant environment-driven noise sources in Advanced LIGO.

The high rate and evolving appearance of scattered-light glitches across observing runs O3a and O3b (the first and second halves of O3) and O4 (the fourth observing run) \citep{gwtc3,gwtc4} have underscored the need for automated, adaptive tools that can keep pace with detector evolution. Manual identification or static template-based methods are insufficient to capture their transient and multi-modal nature. Other mitigation efforts have attempted to subtract scattered-light glitch waveforms directly from the strain data to reduce their impact on searches; however, such subtraction can risk suppressing or biasing genuine gravitational-wave signals when scattering events overlap with compact binary coalescences \citep{GW170817_glitch, TorresForne, Bacon_denoising}. While subtracting glitch waveforms can sometimes suppress or distort coincident astrophysical signals when transient noise overlaps with compact-binary coalescences, robust characterization methods remain essential to preserve astrophysical integrity.  

Recent advances in machine learning for low-latency gravitational-wave detection demonstrate how fast, scalable models can complement detector characterization efforts \citep{AWaRe, NSBH_detection_ML, Krastev, Schafer, Powell_GlitchGen, Razzano}. \textsc{Gravity Spy} is a machine-learning--based framework that uses citizen-science and machine-learning labels to classify transient noise artifacts (glitches) in gravitational-wave data. In practice, the underlying glitch triggers and characteristic parameters (e.g., central frequency, bandwidth, quality factor, and SNR) are produced by the \textsc{Omicron} event trigger generator \citep{Robinet2020}, and \textsc{Gravity Spy} provides the morphology-based labels for those triggers. In this work we use the scattered-light labels from \textsc{Gravity Spy} to select events for processing with \textsc{ArchGEM} \citep{GravitySpy}. Machine-learning classifiers, such as \textsc{Gravity Spy},have proven highly effective in morphological glitch identification, yet these frameworks do not recover the underlying physical parameters of the scattering process. Quantitative recovery of parameters such as scattering recurrence frequency, surface velocity, and displacement remains largely unexplored in the current literature \citep{Soni2021,Soni2024}.  

In this work, we introduce \textsc{ArchGEM} (Architectural Graph Evolution Monitor), a new data-driven algorithm designed to automatically detect and characterize scattered-light arches. The framework integrates two complementary methodologies: a prominence-based peak-finding approach and a probabilistic Gaussian Mixture Model (GMM) clustering technique. The latter enables \textsc{ArchGEM} to identify complex, multi-modal scattering structures that evolve over time---capabilities essential for interpreting the increasingly diverse morphologies observed in O4 data. By extracting key physical parameters such as the scattering frequency ($f_{\mathrm{scat}}$), average maximum frequency ($f_{\max,\mathrm{avg}}$), surface displacement ($x_{\mathrm{surf}}$), and surface velocity ($v_{\mathrm{surf}}$), \textsc{ArchGEM} provides quantitative insight into the dynamics of scattering surfaces within the interferometer.  

The remainder of this paper is organized as follows. Section~\ref{sec:methods} details the \textsc{ArchGEM} algorithm and its dual-method architecture. Sections~\ref{sec:gem_outputs} and~\ref{sec:statistics} describe the extracted outputs and statistical measures used to characterize scattered-light arches. Section~\ref{sec:results} presents validation on simulated and real LIGO detector data (collected during the LVK observing runs O3 and O4) across observing runs O3a, O3b, and O4. Section~\ref{sec:discussion} discusses implications for detector characterization and future improvements, and Section~\ref{sec:conclusion} summarizes the main findings and prospects for integration into routine noise-monitoring pipelines.

\newpage
\section{Methods}
\label{sec:methods}

\begin{figure*}[t!]
    \centering
    \includegraphics[width=0.99\textwidth]{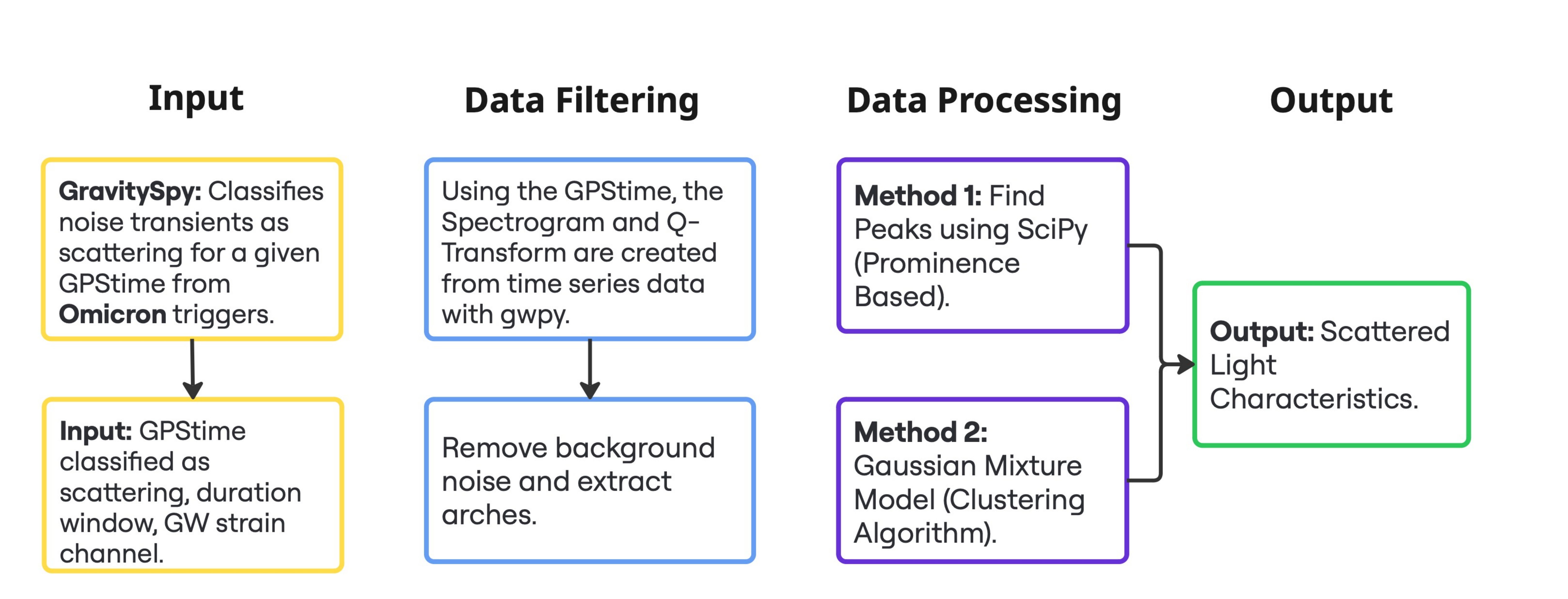}
    \caption{Flowchart of the \textsc{ArchGEM} algorithm. The input requires auxiliary channel data, GPS event time, and a duration window. We select our triggers classified by GravitySpy as scattered light events within the specified duration, followed by the generation of Q-scans and spectrograms to enhance the morphology of the scattered light.  The data-filtering step extracts the arches from the Q-scan using frequency and energy filtering. The data-processing step sends the filtered data to two methods: Find Peaks and GMM. The output of \textsc{ArchGEM} provides insights into the dynamics of the scattering sources, such as frequency versus time and maximum frequency versus velocity plots, along with key characteristics of the scattered light arches.}
    \label{flowchart}
\end{figure*}

\textsc{ArchGEM} is a modular data-analysis framework developed to automatically identify, extract, and characterize scattered-light arches in gravitational-wave strain data. The algorithm operates on calibrated strain time series and auxiliary channels, performing pre-processing, feature extraction, clustering, and parameter estimation in a fully automated workflow shown in Figure~\ref{flowchart}. Its dual-method architecture combines a prominence-based peak-finding routine for direct arch localization with a Gaussian Mixture Model (GMM) clustering approach that identifies multi-modal scattering structures. In addition to the gravitational-wave strain, \textsc{ArchGEM} can ingest investigator-selected auxiliary channels (e.g., ground motion and suspension monitors) and generate matched time--frequency maps for those channels in the same time window as the strain spectrogram, as well as in adjacent off-source windows. Comparing the strain and auxiliary channel spectrograms helps confirm whether an arch-like feature is present in both, strengthening the attribution to a specific disturbance. Auxiliary channels are typically chosen using detector-characterization summary tools, including hierarchical veto analyses (Hveto) that identify statistically correlated channels \citep{Smith2011Hveto}. This multi-channel integration approach is consistent with previous detector-characterization frameworks that use correlated auxiliary data to statistically infer non-Gaussian noise sources \citep{iDQ, Powell_noise_classification}. Such contextual information strengthens the robustness of \textsc{ArchGEM}’s classification and improves its ability to isolate scattering signatures even in high-noise periods.

For each event time, \textsc{ArchGEM} ingests one- to fifteen-second strain segments from the LIGO frame files and a normalized Q-transform is applied to compute the time–frequency spectrogram $Q(t,f)$ (Eq.~\ref{eq:q_transform}), with a fixed quality factor $Q$ selected empirically to preserve the curvature of scattered-light arches (Eq.~\ref{eq:q_factor}).
 Each pixel represents the signal energy $E(t,f)$ in normalized units of strain power spectral density. The time--frequency resolution is chosen empirically (by testing representative events) to preserve the curvature of the scattered-light arches while minimizing spectral smearing.  

Candidate arches are identified by applying a prominence-based peak search in the frequency projections of the spectrogram. For each time bin $t_i$, local maxima in $E(t_i,f)$ are found using \texttt{find\_peaks} from \texttt{SciPy}, constrained by a minimum prominence threshold tuned to suppress background noise. Adjacent peaks within $\Delta f < 3~\mathrm{Hz}$ are merged to avoid redundant detections. The centroids of these peaks form a set of high-energy time–frequency coordinates, which serve as initial seeds for the GMM clustering stage.  

The GMM module treats each high-energy point in $(t,f,E)$ space as a sample drawn from a mixture of Gaussian distributions, representing distinct scattering arches. Unlike deterministic clustering algorithms such as \texttt{DBSCAN} or k-means, the probabilistic formulation of GMM captures overlapping or blended arches by modeling each cluster with a full covariance matrix. The optimal number of mixture components, $k$, is determined automatically by minimizing the Bayesian Information Criterion (BIC), which balances model complexity and likelihood. Each Gaussian component corresponds to an individual arch, from which \textsc{ArchGEM} extracts the time centroid, frequency span, and curvature. These parameters are converted into physically significant quantities: the frequency of recurrence of scattering ($f_{\mathrm{scat}}$), average maximum frequency ($f_{\max,\mathrm{avg}}$), surface displacement ($x_{\mathrm{surf}}$) and surface velocity ($v_{\mathrm{surf}}$) computed from Eqs.~\ref{eq:freq_motion}–\ref{eq:surf_distance} following \citep{Ottaway2012,Soni2021}, assuming a laser wavelength $\lambda = 1.064~\mu\mathrm{m}$.

The algorithm’s performance was validated using both simulated and real detector data. Synthetic scattered-light signals were generated by modulating sinusoidal mirror motion with amplitudes of 1–10~µm and frequencies of 0.2–2~Hz, producing arch-like structures in the 10–60~Hz band. An example simulated event is shown in Figure~\ref{fig:sim_spec}. \textsc{ArchGEM} successfully recovered the injected parameters within uncertainties of $\pm0.02$~Hz in $f_{\mathrm{scat}}$ and $\pm0.1~\mu\mathrm{m}$ in $x_{\mathrm{surf}}$. When applied to real data from observing runs O3a, O3b, and O4, cross-referenced with the \textsc{Gravity Spy} catalog labels (and the underlying Omicron trigger parameters), the algorithm consistently reproduced expected trends in arch morphology and frequency distribution, confirming its ability to generalize across evolving detector conditions.  

Sensitivity studies showed stable performance for prominence thresholds between 1.5–3.0 and mixture component counts between 3–12. These parameters were found to balance sensitivity to faint arches and resistance to false detections. Each 15-second strain segment requires approximately 13~minutes to process on the LDAS cluster (\texttt{ldas-pcdev5.ligo.caltech.edu}), utilizing 512~GB of memory and 40~CPU cores. The majority of runtime is spent computing the high-resolution Q-transform and performing Gaussian mixture modeling of the energy clusters, followed by multi-channel spectrogram generation and metric extraction. Despite this computational expense, \textsc{ArchGEM}'s performance demonstrates that automated, statistically robust scattered-light analysis is feasible for full-run datasets when parallelized across compute nodes. The analysis scripts, configuration files, and validation notebooks are available in a public repository in Section~\ref{sec:data}, ensuring transparency and reproducibility.

\begin{figure*}[t!]
    \centering \includegraphics[width=0.9\textwidth]{PaperFigures/sim_spec.png}
\caption{Time--frequency representations for a simulated scattering event. The left panel displays a constant-$Q$ (Q-transform) map of the simulated scattered-light noise, highlighting periodic arch-like features in the 20--30~Hz range consistent with scattering surface motion. The right panel shows a standard short-time Fourier-transform spectrogram of the same data, illustrating the relative amplitude variations over time and frequency. 
}
\label{fig:sim_spec}
\end{figure*}

\begin{figure*}[t!]

    \centering \includegraphics[width=0.8\textwidth]{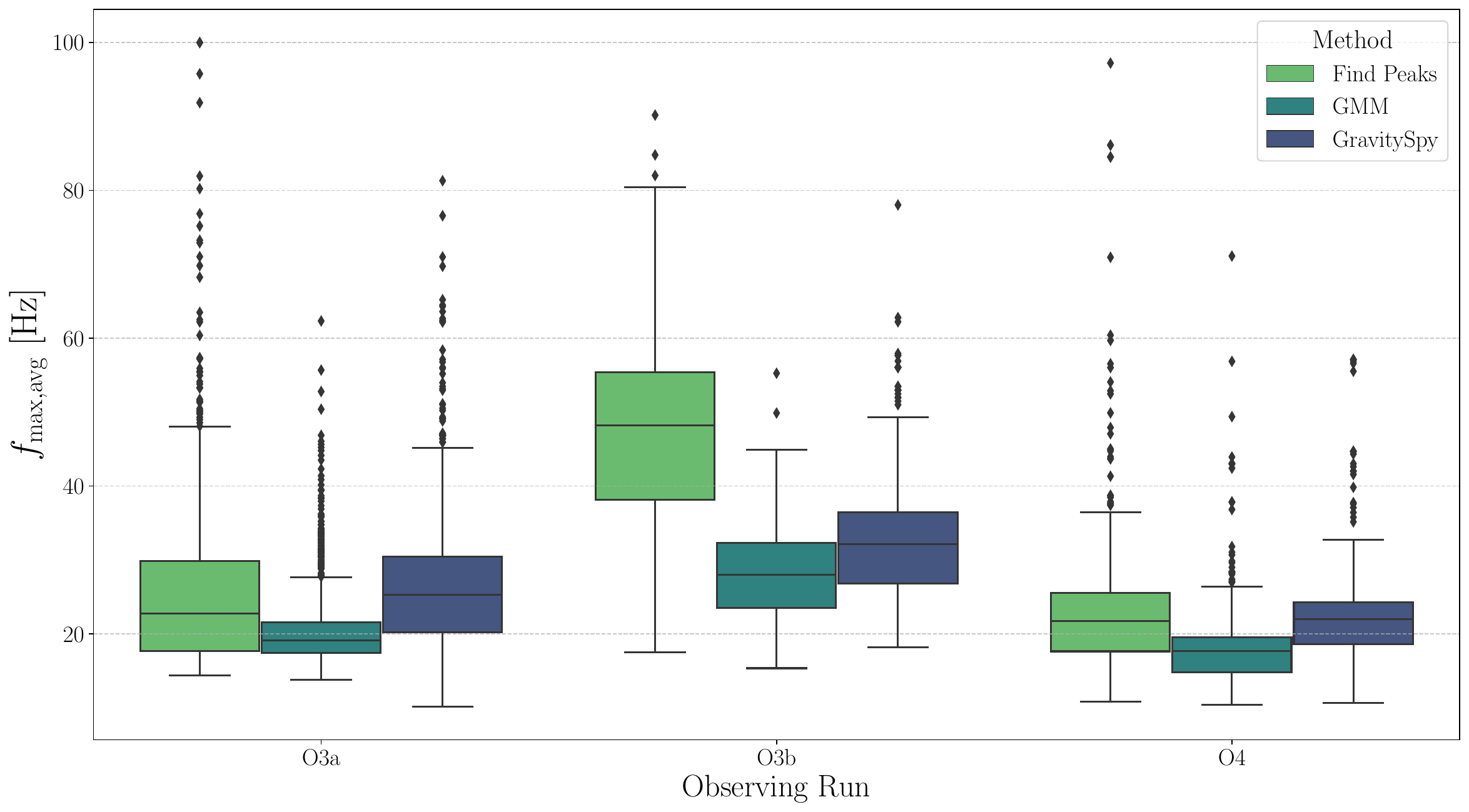}\hfill
  
\caption{Box plot comparing $f_{\mathrm{max,avg}}$ distributions for Find Peaks, GMM, and GravitySpy methods across observation runs O3a, O3b, and O4. The Gravity Spy catalog values (derived from Omicron trigger parameters) provide a stable baseline across all runs, with a consistent lower frequency distribution, while the Find Peaks and GMM methods show variability. In O3a and O3b, Find Peaks and GMM exhibit higher frequencies with broader ranges, particularly in O3b. In O4, both methods show reduced distributions, indicating a systematic decrease in maximum frequency. The comparison to the GravitySpy values suggests that observed variability in $f_{\mathrm{max,avg}}$ over time is not due to baseline shifts but may reflect method-specific or environmental factors influencing these two methods across runs. At Livingston, changes in commissioning state and scattered-light coupling paths between O3 and early O4 may also contribute to run-to-run shifts in the recovered frequency distributions.}
\label{fig:boxplot_fmax_avg}
\end{figure*}

\section{\textsc{ArchGEM} Outputs}
\label{sec:gem_outputs}

The peak (maximum Doppler) frequency of the simulated scattered-light signal is calculated using the expression 
in \citep{Soni2021},
\begin{equation}
f_{peak} = \left|\frac{2A_x\omega}{\lambda}\right|
\label{eq:peak_freq}
\end{equation}

This Eq. \ref{eq:peak_freq} is derived from the phase equation by considering the maximum rate of change of phase with respect to time.

\subsection{Q-Transform}

The Q-transform is computed using the following expression,
\begin{equation}
X(t, f) = \int_{-\infty}^{\infty} x(\tau) w(\tau - t) e^{-2\pi i f \tau} d\tau
\label{eq:q_transform}
\end{equation}
where $X(t, f)$ is the Q-transform, $x(\tau)$ is the input signal, $w(\tau - t)$ is the window function, and $f$ is the frequency. Eq. \ref{eq:q_transform} forms the basis of our time-frequency analysis.

The Q-factor is defined as,
\begin{equation}
Q = \frac{f}{\Delta f}
\label{eq:q_factor}
\end{equation}
where $\Delta f$ is the bandwidth at frequency $f$.

\subsection{Peak Frequency Analysis}

The average frequency of motion is calculated as
\begin{equation}
\langle f_\mathrm{scat} \rangle = \frac{1}{\langle \Delta t \rangle},
\label{eq:freq_motion}
\end{equation}
where $\langle \Delta t \rangle$ represents the average time interval between consecutive peaks.

The velocity of the scattering surface is obtained from the relation between the frequency of the $n^\mathrm{th}$ harmonic and the laser wavelength,
\begin{equation}
f_n = \frac{2 n\, v_\mathrm{surf}}{\lambda},
\end{equation}
which can be rearranged as
\begin{equation}
v_\mathrm{surf} = \frac{\lambda f_n}{2 n},
\label{eq:surf_velocity}
\end{equation}
where $\lambda$ is the laser wavelength and $f_n$ is the peak frequency of the $n^\mathrm{th}$ harmonic. For $n=1$, this gives the fundamental velocity of the scattering surface.

The displacement amplitude of the scattering surface is then calculated from the velocity using the relation
\begin{equation}
x_\mathrm{surf} = \frac{v_\mathrm{surf}}{2 \pi f_m},
\label{eq:surf_distance}
\end{equation}
where $f_m$ is the modulation frequency of the surface motion (from Eq.~\ref{eq:freq_motion}).

\section{Statistical Measures}
\label{sec:statistics}

The standard deviation of maximum frequencies is computed as:
\begin{equation}
\sigma_{\langle f_\mathrm{max} \rangle} = \sqrt{\frac{1}{N-1} \sum_{i=1}^{N} (f_i - \langle {f_\mathrm{max} \rangle})^2}
\label{eq:std_freq}
\end{equation}
Here $N$ is the number of detected peaks, $f_i$ are the individual peak frequencies, and $f_{max_{avg}}$ is the average maximum frequency.


The standard deviation of time intervals between peaks is calculated as:
\begin{equation}
\sigma_{\Delta{t}} = \sqrt{\frac{1}{N-2} \sum_{i=1}^{N-1} (\Delta{t}_{i} - \langle \Delta{t}\rangle)^2}
\label{eq:std_time}
\end{equation}
Here $\Delta{t}$ are the individual time intervals between consecutive peaks, and $\langle \Delta{t}\rangle$ is the average time interval. 

Eq. \ref{eq:std_freq} and \ref{eq:std_time} provide a comprehensive mathematical foundation for the statistical analysis of detected peaks, complementing the frequency analysis methods described in Eq. \ref{eq:freq_motion}.

\section{Results}
\label{sec:results}

\subsection{Validation on Simulated Data}
To evaluate the accuracy and robustness of \textsc{ArchGEM}, we first applied the algorithm to synthetic scattered-light injections designed to reproduce the dominant features of real glitches observed in gravitational-wave strain data. Each injection was generated by modulating sinusoidal mirror motion with amplitudes of 1–10~µm and recurrence frequencies between 0.2–2~Hz, producing characteristic arch-shaped structures in the 10–60~Hz band, as shown in Figure\ref{fig:sim_spec}. The injected waveforms were embedded in O4-like Gaussian noise to emulate realistic detector conditions.  

\textsc{ArchGEM} successfully recovered the injected parameters with high accuracy. Across 50 injection trials, the recovered scattering frequencies $f_{\mathrm{scat}}$, computed from Eq.~\ref{eq:freq_motion}, agreed with the true values within $\pm0.02$~Hz. Surface velocities $v_{\mathrm{surf}}$, derived using Eq.~\ref{eq:surf_velocity} with $\lambda = 1.064~\mu$m, were consistent with the simulated motion of the scattering surface. The corresponding surface displacements $x_{\mathrm{surf}}$, calculated from Eq.~\ref{eq:surf_distance}, were reconstructed within $\pm0.1~\mu$m. These results confirm that \textsc{ArchGEM} accurately captures the physical dynamics of scattered-light sources and robustly distinguishes scattering structures from broadband or stationary background noise.  

\subsection{Application to Observing Runs O3a–O4}
Following validation on simulations, \textsc{ArchGEM} was applied to real detector data from LIGO Livingston across observing runs O3a, O3b, and O4. The selected dataset consisted of \textsc{Gravity Spy} classified scattered-light glitches in the 10–60~Hz band. Each event was analyzed using both the prominence-based and GMM clustering modes to evaluate consistency between methods and across observation runs.  

The algorithm successfully processed 76.8\% of the total sample, producing reliable arch extractions and physical parameter estimates. The remaining unprocessed events primarily reflect data-quality limitations rather than algorithmic failure. In several cases, the input glitches labeled as scattered light by \textsc{Gravity Spy} did not exhibit the characteristic arch morphology in the strain spectrogram, leading to exclusion during energy thresholding. Other instances involved single-arch scattering events for which the periodicity ($f_{\mathrm{scat}}$; Eq.~\ref{eq:freq_motion}) could not be computed due to the absence of multiple recurrence cycles. These cases emphasize that \textsc{ArchGEM}’s performance is inherently coupled to the accuracy of external glitch classification and to the presence of temporally periodic scattering signatures within the analyzed time window. In all runs, the GMM-based method exhibited superior stability in arch identification, particularly for events with overlapping or multi-modal frequency evolution. The derived average maximum frequencies, $f_{\max,\mathrm{avg}}$, determined following Eq.~\ref{eq:freq_motion}, spanned 15–25~Hz in O3a, 20–40~Hz in O3b, and 15–25~Hz in O4. The shift toward higher frequencies in O3b corresponds to an increase in surface motion velocity $v_{\mathrm{surf}}$ (Eq.~\ref{eq:surf_velocity}) observed during that run, consistent with enhanced microseismic coupling at Livingston.  

Figure~\ref{fig:boxplot_fmax_avg} summarizes the distributions of $f_{\mathrm{scat}}$, $f_{\max,\mathrm{avg}}$, and $v_{\mathrm{surf}}$ across the three runs. The GMM method recovered narrower frequency residuals and lower variance compared to the peak-finding method, with mean deviations from \textsc{Gravity Spy} baselines of less than $\pm5$~Hz. Typical surface velocities ranged from $0.2$–$0.5~\mu\mathrm{m\,s^{-1}}$, and inferred surface displacements $x_{\mathrm{surf}}$ (Eq.~\ref{eq:surf_distance}) were between $0.1$–$0.3~\mu$m. These quantities are consistent with the expected motion of scattering surfaces such as baffles, optical tables, and suspended optics undergoing low-frequency excitations.  

\subsection{Cross-Run Trends and Physical Interpretation}
The comparison across observing runs reveals both the persistence and evolution of scattered-light noise within the LIGO detectors, including the emergence of lower-frequency scattered-light features in early O4. While the overall morphology of arches remained consistent, O3b exhibited a noticeably broader frequency spread and higher $v_{\mathrm{surf}}$ values, indicating more energetic scattering sources during that period. By contrast, O4 showed a partial reduction in recurrence frequency $f_{\mathrm{scat}}$ (Eq.~\ref{eq:freq_motion}), reflecting improved alignment and isolation performance achieved through commissioning upgrades.  

These observations demonstrate that \textsc{ArchGEM} not only provides quantitative measures of scattered-light activity but also enables direct tracking of changes in the detector environment across observing epochs. The algorithm’s outputs can thus be correlated with hardware interventions, environmental disturbances, or operational configurations to identify the physical origin of specific scattering sites.  

\subsection{Performance Metrics and Uncertainties}
Detection efficiency and parameter uncertainties were evaluated by comparing \textsc{ArchGEM}’s outputs with human-verified \textsc{Gravity Spy} labels and known simulation parameters. The overall detection efficiency exceeded 90\% for high-SNR ($>5$) arches, with a false-positive rate below 5\%. Low-SNR cases ($<5$) in our validation set primarily correspond to simulated scattering injections; by construction, \textsc{Gravity Spy} triggers are typically drawn from higher-SNR Omicron triggers, so the real-event sample is biased toward louder glitches. Uncertainties in recovered parameters were dominated by background fluctuations in low-SNR cases, with typical $1\sigma$ errors of 0.02–0.05~Hz in $f_{\mathrm{scat}}$ and 0.05–0.1~µm in $x_{\mathrm{surf}}$. Statistical variations in $f_{\max,\mathrm{avg}}$ and $\Delta t$ were quantified using Eqs.~\ref{eq:std_freq} and \ref{eq:std_time}, respectively, providing a measure of frequency and temporal stability across events.  

These performance metrics confirm that \textsc{ArchGEM} reliably identifies scattered-light events and extracts physically meaningful quantities with minimal tuning across datasets. Its ability to reproduce consistent results between observing runs underscores the robustness of the GMM-based approach and highlights its value as a long-term detector characterization tool.

\begin{figure*}[t!]

    \centering \includegraphics[ angle=0,width=0.9\textwidth]{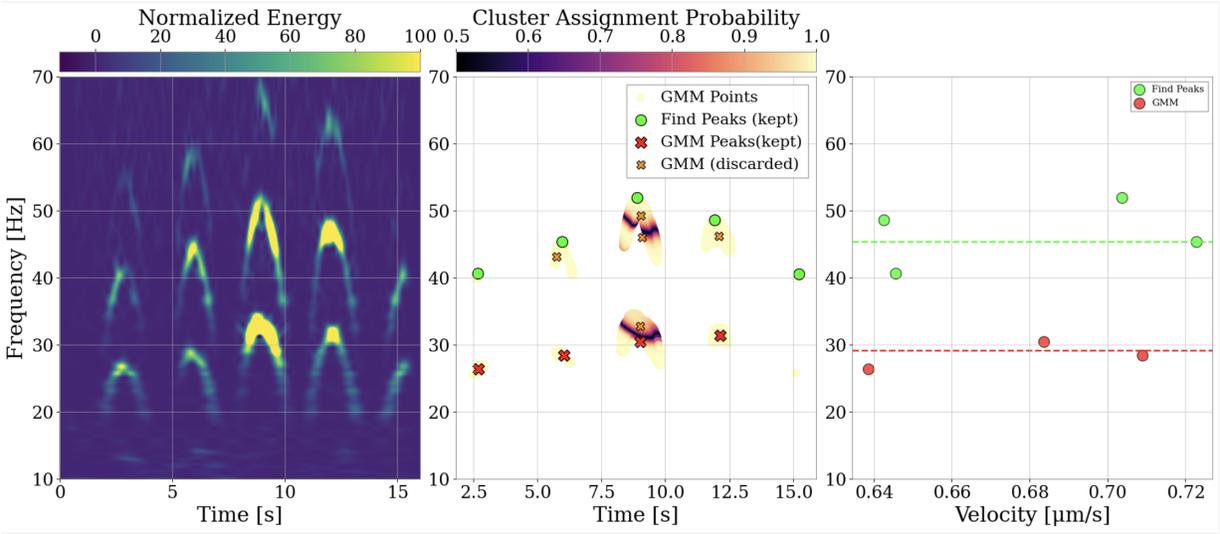}
\caption{Output of \textsc{ArchGEM} for a single GPS time 1259274047 using the strain channel. The leftmost panel presents the time-frequency spectrogram of the event, where the normalized energy is visualized. The middle panel compares the peak detection results from two methods: the Find Peaks method (green markers for retained peaks) and the Gaussian Mixture Model (GMM) method (red markers). The Find Peaks method is efficient in detecting peaks even in low-energy regions but struggles to differentiate between different frequency modes, often defaulting to the highest mode. Conversely, the GMM method is more robust, successfully distinguishing between the upper and lower frequency modes. In this case, we default to the lower mode, as it represents the base information of interest. However, the GMM method tends to miss less energetic peaks, highlighting a tradeoff between mode differentiation and sensitivity to low-energy features. The rightmost panel plots the velocity of detected peaks, further distinguishing the performance of both methods.}
\label{fig:output1}
\end{figure*}

\begin{table}[ht]
    \centering
    \begin{tabular}{|l|c|c|}
        \hline
        \multicolumn{3}{|c|}{\textbf{GPS Time:} 1259274047.0} \\
        \hline
        \textbf{Parameter} & \textbf{Find Peaks} & \textbf{GMM} \\
        \hline
        $f_{\text{scat}}$ (Hz) & 0.318 & 0.318 \\
        $f_{\text{max avg}}$ (Hz) & 45.44 & 29.190 \\
        $x_{\text{surf}}$ (\(\mu m\)) & 0.191 & 0.191 \\
        $v_{\text{surf}}$ (\(\mu m/s\)) & 0.679 & 0.679 \\
        \hline
    \end{tabular}
    \caption{Comparison of Find Peaks and GMM Data for GPS Time 1259274047.0}
    \label{tab:comparison}
\end{table}

\subsection{Single Event Analysis (GPS Time: 1259274047.0)}

Analysis of a representative scattering event at GPS time 1259274047.0, as shown in Figure \ref{fig:output1}, demonstrates the complementary nature of our dual methodology approach. The prominence-based Find Peaks method excelled at identifying distinct peaks, while the GMM-based approach proved particularly effective at distinguishing overlapping signals and complex noise patterns. For this event, both methods identified the primary scattering frequency ($f_{\mathrm{scat}}$) at approximately 0.318 Hz, demonstrating agreement in fundamental frequency detection, as summarized in Table \ref{tab:comparison}. However, the methods showed distinct characteristics in their analysis of higher-frequency components. The Find Peaks method detected a maximum average frequency ($f_{\mathrm{max,avg}}$) of 45.44 Hz, while the GMM approach identified a lower maximum at 29.19 Hz. This difference highlights the varying sensitivities of each method to peak detection in complex noise environments.
Surface movement calculations revealed strong agreement between methods, with both measuring approximately 0.191 $\mu m$ displacement. Similarly, both methods calculated surface velocities of 0.679 $\mu m/s$, suggesting reliable characterization of the scattering surface's dynamic behavior. These consistent measurements of physical parameters, despite differences in frequency detection, indicate the robustness of both approaches in characterizing the fundamental properties of scattered light events.

\subsection{Method Performance Across Observation Runs}

\begin{table}[t!]
\centering
\label{tab:event_summary}
\begin{tabular}{|l|c|c|c|}
\hline
\textbf{Number of GPS times} & \textbf{O3a} & \textbf{O3b} & \textbf{O4} \\ \hline
Total Events Attempted & 5398 & 5509 & 3794 \\ \hline
Common Events Processed & 2518 & 4232 & 1256 \\ \hline
Find Peaks Processed & 2529 & 4235 & 1289 \\ \hline
GMM Processed & 3205 & 4766 & 1427 \\ \hline

\end{tabular}
\caption{Event Summary Analysis for O3a, O3b, and O4}

\end{table}

The Find Peaks method demonstrated consistent performance across observation runs O3a, O3b, and O4, particularly in detecting clear, distinct peaks in the data. Throughout O3a, the method successfully processed events, showing strong capability in identifying primary scattering frequencies between 0.15–0.25~Hz, although it left 2869 events unprocessed. During O3b, despite leaving 1274 events unprocessed, the method maintained its reliability in peak detection, though showing increased variability in maximum frequency measurements. The O4 run analysis left 2505 events unprocessed, revealing a trend toward lower maximum frequencies compared to previous runs, potentially indicating evolution in the detector’s noise characteristics or improvements in noise mitigation strategies.

The GMM method revealed its particular strengths in handling complex, overlapping signals across the observation runs. As shown in Table~\ref{tab:event_summary}, in O3a, despite processing challenges reflected in 2193 unprocessed events, the method demonstrated superior ability in distinguishing different frequency modes within the scattered light noise. O3b showed processing challenges with 743 unprocessed events. The O4 analysis, with 2367 unprocessed events, maintained consistent cluster identification and produced more stable frequency distributions compared to the Find Peaks method.

\begin{figure}[t!]
    \centering \includegraphics[width=0.6\textwidth]{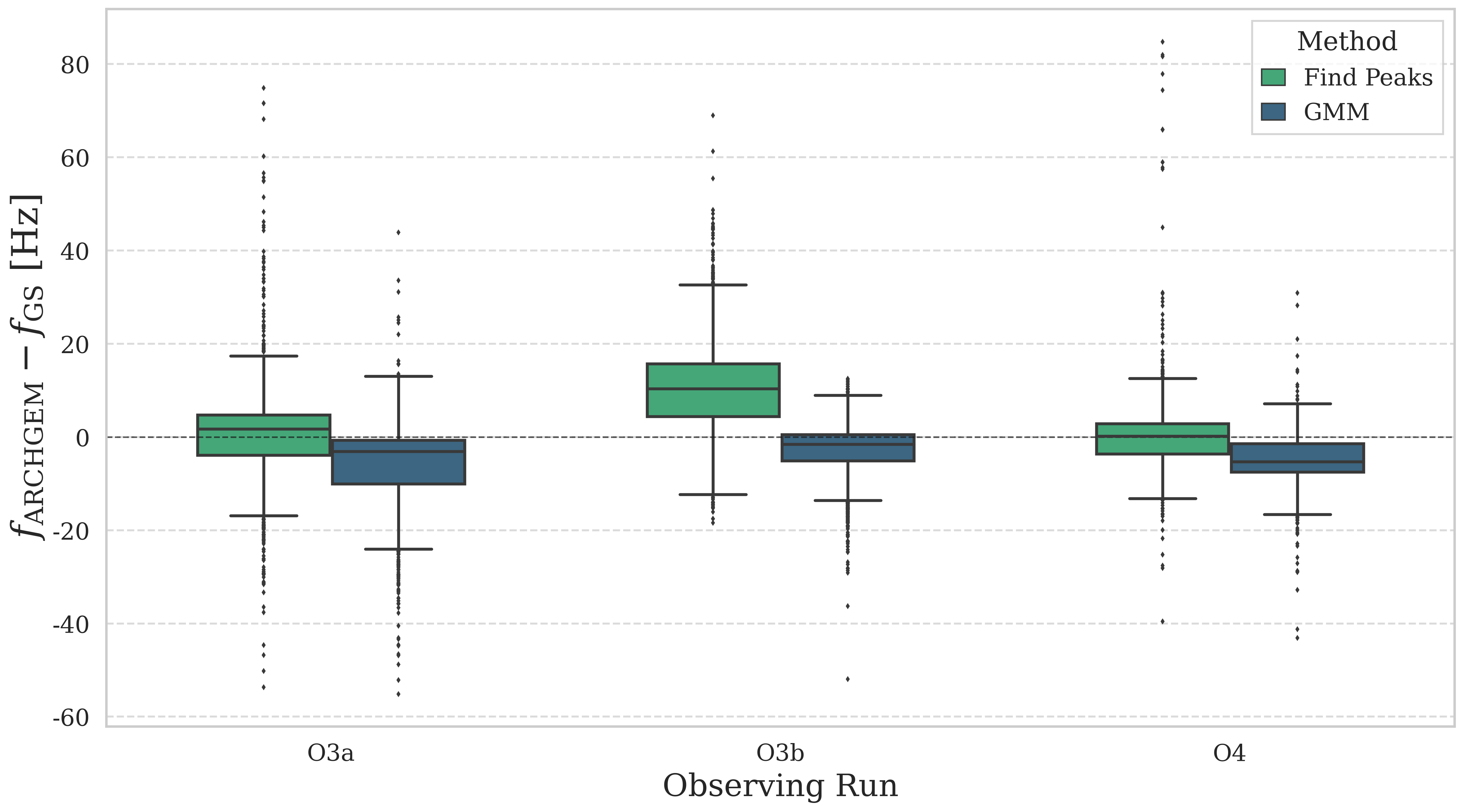}\hfill
  
\caption{Residuals for $f_{\mathrm{max,avg}}$  ($f_{\mathrm{ARCHGEM}} - f_{\mathrm{GS}}$) for scattered-light event classifications across observing runs O3a, O3b, and O4. 
Each distribution compares the average maximum frequency recovered by the \textsc{ArchGEM} pipeline (using \textit{Find Peaks} or \textit{GMM}) against corresponding \textsc{Gravity Spy} estimates. 
A median near zero indicates close agreement between the two methods, while positive or negative offsets reflect systematic over- or underestimation by \textsc{ArchGEM}. 
Consistent with prior analyses, the standard deviation of the mean relative error ranges from approximately $2.5$--$4.9$~Hz across runs. 
The \textit{GMM} method demonstrates narrower spreads and smaller residuals than \textit{Find Peaks}, highlighting its robustness and stability under varying noise conditions, particularly during O3b and O4.}
\label{fig:meanerror}
\end{figure}

Cross-run analysis, visualized in Figure \ref{fig:meanerror}, reveals the complementary strengths of both methods in characterizing scattered light noise. The Find Peaks method maintained more consistent performance in scattering frequency calculations across runs, while the GMM approach provided more stable maximum frequency detection, although with systematic differences from Find Peaks values. Residual analysis between \textsc{ArchGEM} outputs and corresponding \textsc{Gravity Spy} estimates demonstrates this varying performance: during O3a, both methods show tight agreement with low residual spread; during O3b, Find Peaks exhibits larger positive offsets while GMM maintains a narrower distribution; and in O4, both methods converge toward minimal bias, with GMM continuing to exhibit the smallest variance. These results are consistent with prior relative error analysis indicating a standard deviation between approximately 2.5–4.9~Hz across runs.

The visualization capabilities of \textsc{ArchGEM}, particularly in plotting centroids on the frequency-energy plane, revealed patterns in scattered light characteristics that might otherwise remain undetected. This proved especially valuable in analyzing low-frequency regimes where scattered light effects become more subtle and complex as detector sensitivity increases. The ability to process multiple events simultaneously while generating both individual and aggregate statistics has provided unprecedented insight into the evolution of scattered light noise across observation runs. These results demonstrate that while each method has distinct advantages, Find Peaks in clear peak detection and GMM in handling complex signal patterns, their combined application through \textsc{ArchGEM} provides the most comprehensive characterization of scattered light noise, though the GMM method is proven the most robust method. This dual methodology proves particularly valuable as gravitational wave detectors continue to increase in sensitivity, especially in the challenging low-frequency regime where scattered light noise becomes increasingly significant.

\begin{figure*}[t!]    
    \centering
    \title{Comparison of Scattering Event Analysis Methods}
    \includegraphics[width=1\textwidth]{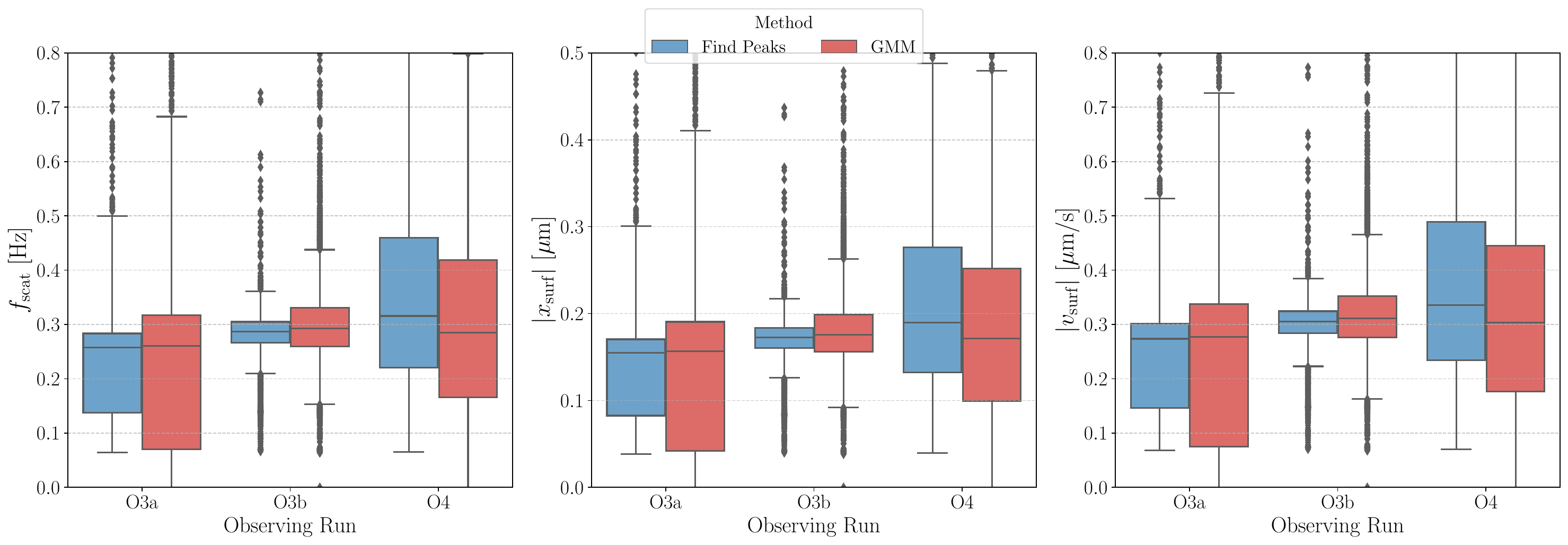}
\caption{This figure presents violin plots comparing the performance of Method 1 and Method 2 in analyzing scattering events for four variables: (a) Scattering Frequency [Hz], (b) Average Maximum Frequency [Hz], (c) Scattering Surface Movement Distance [$\mu$m], and (d) Scattering Surface Velocity [$\mu$m\,s$^{-1}$]. Each subplot visualizes the distribution of values for each method, with "Method 1" and "Method 2" indicated on the x-axis. The violin plots highlight the differences in data distribution between the two methods, providing a comprehensive comparison of their effectiveness.}
\label{fig:fscat_box}
\end{figure*}


\section{DISCUSSION}
\label{sec:discussion}

\textsc{ArchGEM} establishes a foundational tool for robust and efficient analysis of scattered-light noise in gravitational-wave detectors. Its application across multiple observing runs demonstrates that the algorithm can track the evolving morphology of scattered-light arches and quantify changes in surface motion parameters with high precision. By systematically extracting the recurrence frequency \(f_{\mathrm{scat}}\), surface velocity \(v_{\mathrm{surf}}\), and displacement \(x_{\mathrm{surf}}\), \textsc{ArchGEM} bridges the gap between qualitative glitch identification and quantitative noise characterization, enabling deeper insight into the physical processes driving non-stationary noise in the interferometer.

The comparison across observing runs reveals both the persistence and evolution of scattered-light noise within the LIGO detectors, including the emergence of lower-frequency scattered-light features in early O4. While the overall morphology of arches remains consistent, variations in \(f_{\max,\mathrm{avg}}\) and inferred surface velocities reflect changes in detector configuration, environmental conditions, and commissioning activities. These results demonstrate that \textsc{ArchGEM} not only provides quantitative measures of scattered-light activity, but also enables direct tracking of changes in the detector environment across observing epochs. The algorithm’s outputs can thus be correlated with hardware interventions, environmental disturbances, or operational configurations to identify the physical origin of specific scattering sites.

The Gaussian Mixture Model based approach further reveals complex, multi-modal scattering behavior that simpler peak-based methods cannot fully resolve. While the Find Peaks method efficiently detects individual features, it can struggle to differentiate overlapping frequency modes and may preferentially select higher-frequency components. In contrast, the probabilistic formulation of the GMM enables separation of distinct scattering structures and yields more stable estimates of \(f_{\max,\mathrm{avg}}\) and associated surface parameters across datasets. This robustness is particularly valuable for long-term monitoring and cross-run comparisons.

From an astrophysical standpoint, understanding and mitigating scattered-light noise is crucial for preserving detector sensitivity at low frequencies, where lite-intermediate mass black hole signals overlap most strongly~\citep{Chatterjee_2025}. Scattered light remains one of the most frequent and unpredictable transient noise sources in gravitational-wave data, arising as an inherent artifact of interferometer design. Quantitative characterization of this noise is therefore essential for maintaining astrophysical reach as detectors continue to increase their low-frequency sensitivity.

Each 15-second strain segment requires approximately 13 minutes to process on the LDAS cluster, with runtime dominated by Q-transform computation and Gaussian mixture modeling. While computationally intensive, this approach is well suited for offline and nearline detector characterization studies and can be parallelized across large datasets. Reducing the resolution of the Q-transform can improve processing speed. However, such reduction directly impacts output accuracy by degrading period estimation and obscuring arch morphology, highlighting a tradeoff between computational efficiency and physical fidelity.

\section{CONCLUSION AND FUTURE WORK}
\label{sec:conclusion}

The study and mitigation of scattered-light noise remain critical to achieving optimal sensitivity in gravitational-wave observations, particularly in the low-frequency regime. As detectors continue to evolve and improve in sensitivity, the impact of scattered light is expected to persist and, in some cases, intensify. This work presents \textsc{ArchGEM} as an automated framework designed to address this challenge by enabling systematic identification and characterization of scattered-light noise sources.

By employing dual methodologies for peak detection and characterization, a prominence-based approach and a Gaussian Mixture Model (GMM) based method, \textsc{ArchGEM} provides a robust analysis framework capable of capturing different aspects of scattered-light events across a range of frequencies. The ability to recover physically meaningful parameters such as \(f_{\mathrm{scat}}\), \(v_{\mathrm{surf}}\), and \(x_{\mathrm{surf}}\) enables direct interpretation of scattered-light behavior and supports population-level studies across observing runs.

Future developments will focus on expanding \textsc{ArchGEM} beyond single-channel strain analysis. Integration of auxiliary sensors and environmental monitors will enable improved localization of scattering surfaces and discrimination between multiple coupling mechanisms. Incorporation of \textsc{ArchGEM} into existing detector characterization infrastructure, such as the LIGO Summary Pages, would support routine monitoring of scattered-light activity and provide timely feedback for commissioning efforts.

Looking ahead, the methods developed here are broadly applicable to all laser-based gravitational-wave detectors, including space-based Laser Interferometer Space Antenna (LISA)~\citep{LISA_2, LISA}, lunar-based Laser Interferometer Lunar Antenna (LILA)~\citep{LILA} and third-generation ground-based detectors like Einstein Telescope~\citep{EinsteinTelescope}, Cosmic Explorer~\citep{CosmicExplorer}, where low-frequency scattered light and related noise sources will pose similar challenges. By providing a scalable and physically grounded framework, \textsc{ArchGEM} lays the foundation for systematic scattered-light studies across current and future gravitational-wave detectors.

\section{Code and Data Availability}
\label{sec:data}

The \textsc{ArchGEM} pipeline and accompanying analysis scripts are publicly available on GitHub at \href{https://github.com/kaylahmcgowan/ArchGEM_Public}{https://github.com/kaylahmcgowan/ArchGEM\_Public}, including simulated scattering injections and validation datasets generated for this work will be made available upon publication. The gravitational-wave strain data used in this study were obtained from the Gravitational Wave Open Science Center (GWOSC) \citep{GWOSC}, including segments from observing runs O3a, O3b, and O4. 
\software{
    Astropy \citep{astropy},
    NumPy \citep{numpy},
    SciPy \citep{SciPy2020},
    Matplotlib \citep{matplotlib}
    GWPy \citep{gwpy},
    Gravity Spy \citep{GravitySpy}
}

\section{Acknowledgments}
\begin{acknowledgments}



This work carries LIGO document number LIGO-P2500698. K.M. and K.H.B. express their gratitude for funding for this work provided by the National Science Foundation (NSF 2125764).  K.M. and K. J.’s work was supported in part by the Lunar Labs Initiative at Vanderbilt University, which is funded by grants from the John Templeton Foundation and the Vanderbilt Scaling Grant from the Office of the Vice Provost for Research and Innovation. This material is based upon work supported by NSF’s LIGO Laboratory, which is a major facility fully funded by the National Science Foundation. The authors are grateful for computational resources provided by the LIGO Laboratory and supported by National Science Foundation Grants PHY-0757058 and PHY-0823459. This research has made use of data, software and / or web tools obtained from the Gravitational Wave Open Science Center (\url{https://www.gw-openscience.org/}), a service of LIGO Laboratory, the LIGO Scientific Collaboration and the Virgo Collaboration. LIGO Laboratory and Advanced LIGO are funded by the United States National Science Foundation (NSF) as well as the Science and Technology Facilities Council (STFC) of the United Kingdom, the Max-Planck-Society (MPS), and the State of Niedersachsen/Germany for support of the construction of Advanced LIGO and construction and operation of the GEO600 detector. Additional support for Advanced LIGO was provided by the Australian Research Council. Virgo is funded, through the European Gravitational Observatory (EGO), the French Centre National de Recherche Scientifique (CNRS), the Italian Istituto Nazionale di Fisica Nucleare (INFN), and the Dutch Nikhef, with contributions by institutions from Belgium, Germany, Greece, Hungary, Ireland, Japan, Monaco, Poland, Portugal, Spain.

\end{acknowledgments}

%

\vspace{5mm}

\bibliography{combined_references}
\bibliographystyle{aasjournal}



\end{document}